\documentclass[a4paper,twocolumn]{article}

\title{On the Vapnik-Chervonenkis dimension of the Ising-perceptron}

\author{S.~Mertens\thanks{email: stephan.mertens@physik.uni-magdeburg.de}\\[1ex]
        \small Institut f\"ur Theoretische Physik\\
				\small Otto-von-Guericke Universit\"at, Postfach 4120, D-39016 Magdeburg, Germany
			 }
			 
\usepackage{amstex,graphicx,fancyheadings}

\begin{document}

\maketitle

{\it
{\bf Abstract.} 
The VC dimension of the Ising perceptron with binary patterns is
calculated by numerical enumerations for system sizes $N \leq 31$.
It is significantly larger than $\frac N2$. The data suggest
that there is probably no well defined asymptotic behaviour for
$N\to\infty$.}

\vspace{0.5cm}
The Vapnik-Chervonenkis(VC)-dimension is one of the central quantities
used in both mathematical statistics and computer science to characterize
the performance of classifier systems
\cite{vapnik:chervonenkis:71,vapnik:82}.
In the case of feed-forward neural networks it establishes connections
between the storage and generalization abilities of these systems
\cite{haussler:etal:91,parrondo:vdbroeck:93,engel:94}. In this letter
we will discuss the VC dimension of the Ising-Perceptron with binary patterns.

The VC-dimension $d_{VC}$ is defined via the growth function $\Delta(p)$.
Consider a set of instances $x$ and a system $C$ of binary classifiers $c$:
$x\mapsto\{-1,1\}$ that group all $x\in X$ into two classes labeled by $1$ and
$-1$ respectively. For any set $\{x^\mu\}$ of $p$ different instances
$x^1,\ldots,x^p$ we determine the number $\Delta(x^1,\ldots,x^p)$ of different
classifications $c(x^1),\ldots,c(x^p)$ that can be induced by running
through all classifiers $c\in C$. A set of instances is called {\em shattered} by
the system $C$ if $\Delta(x^1,\ldots,x^p)$ equals $2^p$,
the maximum possible number of different binary classifications of $p$
instances. Large values of $\Delta(x^1,\ldots,x^p)$ roughly correspond to a
large diversity of mappings contained in $C$. The growth function
$\Delta(p)$ is now defined by
\begin{equation}
  \label{growth}
  \Delta(p) = \max_{x^\mu}\Delta(x^1,\ldots,x^p).
\end{equation}

It is obvious that $\Delta(p)$ cannot decrease with $p$. Moreover for small
$p$ one expects that there is at least one shattered set of size $p$ and hence
$\Delta(p) = 2^p$. On the other hand this exponential increase of the growth
function is unlikely to continue for all $p$. The value of $p$ where it starts
to slow down gives a hint on the complexity of the system $C$.
In fact the Sauer lemma \cite{vapnik:chervonenkis:71,sauer:72}
states that for all systems $C$ of binary classifiers there exists a
natural number $d_{VC}$ (which may be infinite) such that
\begin{equation}
  \Delta(p) \left\{
  \begin{array}{lcr}
    = 2^p & \quad\mbox{if}\quad & p \leq d_{VC} \\
    \leq \sum\limits_{i=0}^{d_{VC}} {p \choose i}
    & \mbox{if} & p \geq d_{VC}
  \end{array}
  \right..
\end{equation}
$d_{VC}$ is called the VC-dimension of the system $C$. Note that it will in general
depend on the set $X$ of instances to be classified.

A concrete example for a system of classifiers is given by the well known
perceptrons defined by
\begin{equation}
  \label{perceptron}
  \sigma = \mbox{sign}(\sum_{i=1}^N J_i\xi_i)
\end{equation}
where the weights $\vec{J}\in\Bbb{R}^N$ 
parameterize the perceptron and $\vec{\xi}\in\Bbb{R}^N$ is an instance or pattern
to be classified.
The multiplication of $\vec{J}$ by a constant factor
does not affect the output $\sigma$, so the weights are usually restricted by
$\vec{J}^2 = N$. For this {\em spherical perceptron} the exact result
$d_{VC} = N$ has been obtained analytically \cite{cover:65}. 

The {\em Ising-perceptron} is a spherical perceptron with the additional
constraint $J_i=\pm1$ on the weigths. For real valued patterns $\vec{\xi}\in\Bbb{R}^N$
this constraint does not affect the VC-dimension, i.e.\
$d_{VC} = N$ still holds \cite{mertens:engel:96}.

Since much of the interest in neural networks with
discrete weights is due to their easy technical implementation it is
important to consider not only binary weigths but also binary
patterns $\xi_i=\pm1$. To avoid problems with the
$\mbox{sign}$-function if $\vec{J}\cdot\vec{\xi}$ happens to be
$0$, one introduces a threshold $\Theta = \pm1$ for $N$ even:
$\sigma = \mbox{sign}(\vec{J}\cdot\vec{\xi}+\Theta)$.
Since the VC-dimension for
the Ising-perceptron with $N = 2n$ and $\Theta=\pm1$ is the same as for
$N = 2n+1$ without threshold, we will consider only odd values of $N$
throughout this paper.

The determination of the VC-dimension of the
Ising-perceptron with binary patterns is a hard problem.
Analytical calculations based on the replica
method \cite{engel:weigt:96} are not very helpful since this method is suited
to calculate {\em typical} or {\em average} quantities whereas
the VC-dimension is an extremal concept due
to the $\max$ in eq.~\ref{growth}.
For the spherical perceptron this difference does not really matter, but for
networks with discrete weights it is crucial \cite{mertens:engel:96}.

To get at least a lower bound for $d_{VC}$ it suffices to find a large
shattered set by a smart guess.
Consider the set ($N$ odd):
\begin{eqnarray}
\label{shattered_2}
\vec{\xi}^{(0)} & = & (-1, -1, \ldots, -1, -1) \nonumber\\
\vec{\xi}^{(1)} & = & (-1, -1, \ldots, -1, +1) \nonumber\\
\vec{\xi}^{(2)} & = & (-1, -1, \ldots, +1, -1) \\
                & \vdots & \nonumber\\
\vec{\xi}^{(\frac{N+1}2)} & = & (-1, \ldots, -1, +1, -1, \ldots, -1).\nonumber
\end{eqnarray}
Let 
$\vec{\sigma} = (\sigma_0,\ldots,\sigma_{\frac{N+1}2})$
be an arbitrary output vector. To see how $\vec{\sigma}$ can be realized by
the binary perceptron, we have to distinguish two cases:

{\em First case:} $\vec{\sigma} = (\sigma_0, \sigma, \ldots, \sigma)$ i.e.\
the output values for all patterns except $\vec{\xi}^{(0)}$ are the same.
This output can be realized by the weights
\begin{equation}
  \vec{J} = (-\sigma, \underbrace{\sigma_0, \ldots, \sigma_0}_{\frac{N-3}2},
	          \underbrace{-\sigma_0, \ldots, -\sigma_0}_{\frac{N+1}2}).
\end{equation}

{\em Second case:} For all output vectors different from the first case, we
can assert
\begin{equation}
\label{assertion}
\left|\sum_{i=1}^{\frac{N+1}2}\sigma_i\right| \leq \frac{N-3}2
\end{equation}
since at least one $\sigma_i$ in the sum differs from the rest.
As weights we choose 
\begin{equation}
  \vec{J} = (-\sigma,  k_1, \ldots, k_{\frac{N-3}2},
	          \sigma_{\frac{N+1}2}, \ldots, \sigma_0)
\end{equation}
where $\vec{k}$ can be any $\pm1$-vector with 
\begin{equation*}
  \sum_{i=1}^{\frac{N-3}2} k_i = -\sum_{i=1}^{\frac{N+1}2}\sigma_i.
\end{equation*}
According to eq.~\ref{assertion}, such a vector can always be found.
Again we have $\mbox{sign}(\vec{J}\cdot\vec{\xi}^{(\mu)}) = \sigma_\mu$ for
$\mu = 0, \ldots, \frac{(N+1)}2$.

This proves that the set (\ref{shattered_2}) is shattered and hence
\begin{equation}
  \label{first_bound}
	d_{VC} \geq \frac{N+3}2
\end{equation}
for the Ising-perceptron with binary patterns. This value of $d_{VC}$ agrees
very well with numerical results obtained by a statistical enumeration
method \cite{stambke:92,mertens:engel:96}. For this method, one {\em randomly}
draws $p$ binary patterns and calculates
$\Delta(\vec{\xi}^{(1)},\ldots,\vec{\xi}^{(p)})$ by enumeration of all
perceptrons $\vec{J}\in\{\pm1\}^N$. If a single pattern set with
$\Delta(\ldots)=2^p$ is found, we know that $d_{VC}\geq p$.
Like the replica method, this method is not suited to calculate 
the VC dimension in cases
where the maximum shattered sets are rare.

There is however a method that guarantees the exact evaluation
of the VC dimension: {\em exhaustive enumeration} of all
shattered sets. The shattered pattern sets can be arranged as the nodes of a tree.
The root of the tree is the empty pattern set (conveniently defined to be
shattered). The children of a $P$ pattern node are formed by
all those shattered $(P+1)$-pattern sets that can be obtained from the
parent by adding a new pattern. The recursive application of this definition
gives the complete tree of all shattered sets. The VC dimension is the height
of the tree. It can be measured by a traversal of the complete tree using standard algorithms.
 
The branching factor of the tree is $O(2^N)$, its height is $O(N)$, 
giving an overall complexity of $O(2^{N^2})$. This exponential complexity limits
the reachable size $N$ very soon and calls for some tricks to reduce the
number of nodes.

Before we can think of reducing the number of nodes, we must ensure
that every node, i.e.\ every shattered set is considered only once.
A $\pm1$-pattern can be read as an $N$-bit integer (identifying
$-1$ with $0$), hence we have an {\em order relation} among the patterns.
If we add only patterns to a set which are larger than the current
elements of the sets, uniqueness of the nodes is guaranteed.

The first trick to reduce the number of nodes exploits the symmetrie
of the problem: A shattered set
remains shattered if we multiply one of its elements or the $i$-th entry
of all elements by $-1$.
Therefore we may restrict ourselves to pattern sets where all elements
start with $-1$: $\vec{\xi}=(-1,\ldots)$ and we can fix the set containing
the only pattern $(-1,-1,\ldots,-1)$ as the root of the tree.

The second trick is of the {\em branch and bound} variety and exploits the
fact, that we are not interested in the complete tree but only in its height.
Let us assume that we have an easy to calculate
upper bound for the maximum height that can be reached from a
given node. If this upper bound turns out to be lower than the
maximum height already found during our traversal,
we can safely renounce the exploration of the subtree rooted in this
node!

The binary outputs
of a set of $P$ patterns can be interpreted as $P$-bit number $c$. 
Iterating over all $2^N$ binary weight vectors of our network,
we get $2^N$ such output numbers $c$. If $P<N$, some of the $c$ values
must appear more than once. Let $f_c$ denote the frequency of the output
value $c$.
The number of different classifications of this pattern set is given
by the number of $f_c > 0$:
\begin{equation}
  \Delta(\xi^1,\ldots,\xi^P) = \sum_{c=0}^{2^P-1} \Theta(f_c).
\end{equation}
The $f_c$'s have to be calculated at each node to test whether the
pattern set is shattered or not. If
\begin{equation}
  f_{\min} = \min_c\{f_c\}
\end{equation}
is 0, the pattern set is not shattered (at least one classification $c$
has not been realized). If $f_{\min}>0$ the pattern set is shattered and
we can try to enhance it. Each new pattern can split an existing
classification into two (appending a $-1$ to $c$ for some weight vectors
and a $+1$ for others), i.e.\ from each classification $c$ we get 2
new classifications $c_1$ and $c_2$ with $f_c = f_{c_1} + f_{c_2}$.
One of the new frequencies is always $\leq f_c/2$.
Therefore we have $\log_2f_{\min}$
as an upper bound for the number of patterns
that can be added to a shattered set before we definitely get a
non shattered set.

This strategy allows to prune many subtrees. For $N=5$, branch
and bound reduces the number of nodes from 77 to 4, for $N=7$
from $8389$ to $4625$.

Even with these tricks, the complexity $O(2^{N^2})$ is overwhelming.
On an UltraSparc I 170, the exhaustive enumeration for $N=7$ takes
less than a second. For $N=9$, the running time is $6.5$ hours!
Nevertheless, the results obtained for $N\leq 9$ are already
quite remarkable. For $N=7$, the set
\begin{eqnarray}
\label{shattered_3}
\vec{\xi}^{(1)} & = & (-1,-1 -1,+1,+1,+1,+1) \nonumber\\
\vec{\xi}^{(2)} & = & (-1,+1 +1,-1,-1,+1,+1) \nonumber\\
\vec{\xi}^{(3)} & = & (-1,+1 +1,+1,+1,-1,-1) \nonumber\\
\vec{\xi}^{(4)} & = & (+1,-1 +1,-1,+1,-1,+1) \\
\vec{\xi}^{(5)} & = & (+1,-1 +1,+1,-1,+1,-1) \nonumber\\
\vec{\xi}^{(6)} & = & (+1,+1 -1,-1,+1,+1,-1) \nonumber\\
\vec{\xi}^{(7)} & = & (+1,+1 -1,+1,-1,-1,+1) \nonumber
\end{eqnarray}
is shattered, hence $d_{VC} = 7$ -- the maximum possible value!
Together with $d_{VC}=4$ for $N=5$ and $d_{VC}=7$ for $N=9$, these results
do not allow a decent conjecture for the general expression.
However, partial enumerations for larger values of $N$ indicate, that
$d_{VC}$ is substantially larger than the value $\frac{N+3}2$ provided
by (\ref{shattered_2}).

The largest shattered sets found by exhaustive and partial enumerations
share a common feature: They can be transformed into quasi-orthogonal sets,
i.e.\ into sets, where the patterns have minimum pairwise 
overlap\footnote{Exact orthogonality cannot be achieved for $N$ odd.},
\begin{equation}
  \label{orthogonality}
  \vec{\xi}^{(\mu)}\cdot\vec{\xi}^{(\nu)} = \left\{
  \begin{array}{rr}
     \pm 1 & \mu\neq\nu\\
		 N & \mu=\nu
  \end{array}
  \right..
\end{equation}
This observation leads to the idea of restricting the enumeration to 
quasi orthogonal pattern sets.

To find such pattern sets, the notion of {\em Hadamard-matrices}
is useful (see e.g.\ \cite{krisement:90} or any texbook on combinatorics or
coding theory). A Hadamard matrix is an
$m\times m$-matrix $H$ with $\pm1$-entries such that 
\begin{equation}
  \label{hadamard}
  HH^T = mI
\end{equation}
where $I$ is the $m\times m$ identity matrix. 
The rows (or columns) of a Hadamard matrix form a 
set of $m$ orthogonal
binary patterns! This implies that $m$ must be even, but the whole truth is
more restrictive: If $H$ is an $m\times m$ Hadamard matrix, then $m=1$, $m=2$
or $m\equiv0\bmod4$. The reversal is a famous open question: Is there a Hadamard
matrix of order $m=4n$ for every positive $n$? The first open case is
$m=428$. 

For special values of $m$ there are rules to construct Hadamard
matrices \cite{beth:etal:85}, e.g.:
\begin{itemize}
\item $m = 2^n$ (Sylvester type)
\item $m = q+1$ where $q$ is a prime power and $q\equiv3\bmod4$ (Paley type).
\item $m = 2(q+1)$ where $q$ is a prime power and $q\equiv3\bmod4$ (Paley
type).
\end{itemize}
These rules provide us with Hadamard matrices of
sufficient size\footnote{The first value of $m = 4n$ where none of them applies
is $m = 92$.}. To get from a $4n\times4n$ Hadamard matrix to quasi orthogonal binary
patterns we either cut out one column ($N = 4n-1$) or add an arbitrary
column ($N = 4n+1$) and take the rows of the resulting matrix as patterns.
The pattern set~(\ref{shattered_3}) is a result of this procedure applied
to the $8\times8$ Hadamard matrix $H_8$ of Sylvester type:
\begin{equation}
 H_8 = H_2 \otimes H_2 \otimes H_2
\end{equation}
with
\begin{equation}
 H_2 = \left(
\begin{array}{cc}
  -1 & -1 \\
	-1 & +1
\end{array}
\right).
\end{equation}
$\otimes$ denotes the usual Kronecker product.

The restriction to quasi orthogonal pattern sets allows us to consider
larger values of $N$, but now the enumeration gives only lower bounds
for $d_{VC}$.
\begin{figure}[htb]
  \includegraphics[scale=.4]{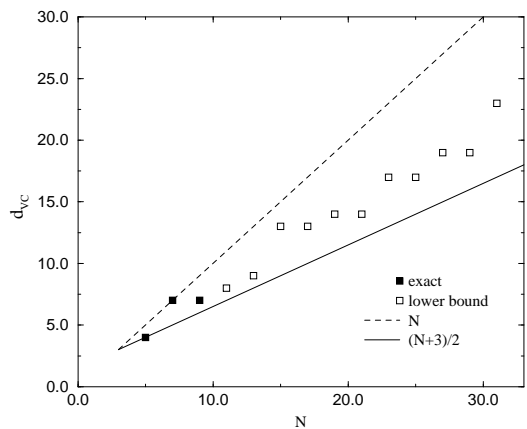}
 	\caption[Fig1]{\label{hadamard_fig}VC-dimension of the Ising perceptron with binary patterns
	vs.\ $N$.
	$d_{VC} = N$
	is an upper bound, $d_{VC} = \frac{N+3}2$ is a lower bound provided by the
  set~(\ref{shattered_2}).}
\end{figure}
Results for $N\leq31$ are displayed in figure \ref{hadamard_fig}.
The lower bound $d_{VC} = \frac{N+3}2$ achieved by the set~(\ref{shattered_2})
is exceeded for all $N > 5$, but the theoretic upper bound  $d_{VC} = N$ is
attained only for $N = 7$. The data are not suited for a decent conjecture
about a general expression for $d_{VC}(N)$. Even the mere existence of a well
defined asymptotic behaviour for $N\to\infty$ looks questionable.
The VC dimension seems to be sensitive not only to the size but also
to the numbertheoretic properties of $N$: We observe a jump in $d_{VC}(N)$
at $N = 2^n - 1$, i.e.\ at values of $N$ where the corresponding Hadamard matrix is of
Sylvester type.

The lower bounds in figure \ref{hadamard_fig} do not rule out the
possibility of a much more regular behaviour of the true $d_{VC}(N)$,
including well defined asymptotics. However, if the limit
$\lim_{N\to\infty}\frac{d_{VC}}N$ exists, it will be probably larger than
$0.5$.

{\em Acknowledgements.} The author appreciates fruitful discussions with
A.~Engel. Thanks to C.~Bessenrodt for her reference to Hadamard-matrices.

\bibliographystyle{unsrt}
\bibliography{vc}

\end{document}